\documentclass[12pt,epsf]{article}

\begin{document}

\begin{titlepage}

\begin{flushright}
\end{flushright}
\vskip 2.5cm

\begin{center}
{\Large \bf Exact Normal Modes for a System with Three Luttinger Liquids
Interacting in an Elongated Trap}
\end{center}

\vspace{1ex}

\begin{center}
{\large B. Altschul\footnote{{\tt baltschu@indiana.edu}}}

\vspace{5mm}
{\sl Department of Physics} \\
{\sl Indiana University} \\
{\sl Bloomington, IN 47405 USA} \\

\end{center}

\vspace{2.5ex}

\medskip

\centerline {\bf Abstract}

\bigskip

We consider a system of spinless fermions in a nearly one-dimensional
cylindrical trap, with the Fermi level such that only the lowest-lying states
with angular
momenta $\ell=+\hbar$, 0, or $-\hbar$ about the axis of the trap are occupied.
We treat the particles in these states as comprising three separate Luttinger
liquids, with the possibility for forward and backward scattering between them.
We determine the normal modes in the system in the presence of this
scattering, finding two sets of linear harmonic fluid modes and one set of
modes that are self-coupled through a sine-Gordon interaction.

\bigskip 

\end{titlepage}

\newpage

The problem of spinless fermionic particles trapped in a cylindrical
potential with a very high aspect ratio possesses many interesting aspects.
If the fermions are constrained to move almost exclusively along the direction
of the traps' axis, then the system becomes effectively one-dimensional in
character. This will affect the qualitative nature of the system's
behavior, since fermions in one dimension have many
curious properties.

In particular, a one-dimensional system of low-temperature fermions
exhibits bosonization properties~\cite{ref-schulz,ref-solyom,ref-emery1}, as
the coherent fermion-hole excitations behave like bosons. This phenomenon is
described by the Luttinger model. The main physical applications of
this model occur in the context of the quantum Hall
effect~\cite{ref-wen1,ref-wen2}. However, the model is clearly also relevant
to a trapped fermion system such as we have described. The field of fermion
trapping is advancing very
rapidly~\cite{ref-demarco,ref-schreck,ref-granade,ref-hadz}, and it may be
possible in the forseeable future to observe Luttinger-liquid-type
behavior in trapped fermionic gasses.

We consider a cylindrical trap with hard walls, of length $L$ and axial radius
$b\ll L$. The fermion modes in this trap are described by
three quantum numbers---$k$, $\ell$, and $n_{\rho}$---corresponding to the
cylindrical coordinates $(z,\theta,\rho)$. $k$ is the momentum along the axis
of the trap (setting $\hbar=1$), while $\ell$ is the angular momentum about
this axis, and $n_{\rho}$ is the number of radial nodes in the wavefunction.
The wavefunction is periodic the $z$-direction with period $L$ and
vanishes at $\rho=b$. In the absence of other interactions, the energy of a
trapped particle is
\begin{equation}
E_{k,\ell,n_{\rho}}=\frac{1}{2m}\left[k^{2}+\left(\frac{j_{\ell,n_{\rho}}}{b}
\right)^{2}\right],
\end{equation}
where $m$ is the fermion mass and
$j_{\ell,n_{\rho}}$ is the $n_{\rho}$-th root of the Bessel function
$J_{\ell}$. For fixed $k$, the three lowest-lying levels have $n_{\rho}=0$ and
$\ell=+1$, 0, and $-1$. The energy gap for the two $|\ell|=1$ states is
$\Delta=\frac{(j_{1,1})^{2}-(j_{0,1})^{2}}{2mb^{2}}\approx\frac{4.44}
{mb^{2}}$.

Previously, we have considered the effects of virtual particles in radially and
angularly
excited states of the trap on a Luttinger liquid lying entirely in the ground
state~\cite{ref-altschul}. In this paper, we consider a different regime, in
which the Fermi energy $\epsilon_{F}$ is larger than $\Delta$, so that the
lowest-lying $\ell=\pm1$ states are filled with large numbers of particles,
while all the higher excited states of the trap remain empty. We shall
treat this system as consisting of three interacting
Luttinger liquids and shall study the normal modes of excitation. For
particular
choices of couplings, there can exist extremely simple exact solutions of this
model.

In accordance with the Luttinger model~\cite{ref-luttinger,ref-mattis1},
we linearize the fermion spectrum and append an infinite negative energy
Dirac sea to the physical system. If we fix the number of particles with each
value of $\ell$, then
we can take the noninteracting Hamiltonian (ignoring any
additive constant) to be
\begin{equation}
\label{eq-h0}
H_{0}=\sum_{\ell=-1}^{+1}\left\{\frac{2\pi v^{\ell}_{F}}{L}\sum_{q>0}\left[
\rho^{\ell}_{+}(q)\rho^{\ell}_{+}(-q)+\rho^{\ell}_{-}(-q)\rho^{\ell}_{-}(q)
\right]\right\},
\end{equation}
where the $\rho^{\ell}_{\pm}$ operators are
\begin{eqnarray}
\label{eq-bosonization}
\rho^{\ell}_{+}(q) & = & \sum_{k}a^{\ell\dag}_{k+q}a^{\ell}_{k}, \\
\rho^{\ell}_{-}(q) & = & \sum_{k}b^{\ell\dag}_{k+q}b^{\ell}_{k}.
\end{eqnarray}
The $a^{\ell}$ ($b^{\ell}$) operators correspond to right- (left-)moving
(spinless) fermions of angular momentum $\ell$ about the axis of the trap, and
$v^{\ell}_{F}$ is the Fermi velocity for the Luttinger liquid of angular
momentum $\ell$. (For quantities, such as the Fermi velocity, which depend only
upon $|\ell|$, we shall never write a negative superscript; only ``0'' or ``1''
will be used.)
For $\ell=0$, we have $v^{0}_{F}=\sqrt{\frac{2\epsilon_{F}}{m}}$, while
$v^{1}_{F}=\sqrt{\frac{2(\epsilon_{F}-\Delta)}{m}}$.
The $\rho^{\ell}_{\pm}$ obey the Bose commutation relations
$[\rho^{\ell}_{+}(-q),\rho^{\ell'}_{+}(q')]=[\rho^{\ell}_{-}(q),\rho^{\ell'}
_{-}(-q')]=\delta_{qq'}\delta_{\ell\ell'}\frac{qL}{2\pi}$ and $[\rho^{\ell}_{+}
(q),\rho^{\ell'}_{-}(q')]=0$, so
they create and annihilate phonons with energies $|q|v_{F}^{\ell}$.

We may also introduce fermion-fermion scattering terms which are bilinear in the
$\rho^{\ell}_{\pm}$ operators. At low temperatures, nearly all the scattered
particles lie near the Fermi surface, with momenta $k^{\ell}_{F}=mv^{\ell}
_{F}$. We shall consider both forward scattering interactions of the types
$(k^{\ell}_{F};k^{\ell'}_{F})\rightarrow(k^{\ell}_{F};k^{\ell'}_{F})$ or
$(k^{\ell}_{F};-k^{\ell'}_{F})\rightarrow(k^{\ell}_{F};-k^{\ell'}_{F})$ and
backward scattering
$(k^{\ell}_{F};-k^{\ell'}_{F})\rightarrow(-k^{\ell}_{F};k^{\ell'}_{F})$. The
backward scattering is forbidden by momentum conservation unless $\ell=\pm
\ell'$, and it is truly distinct from the forward scattering only if $\ell\neq
\ell'$. [``Inelastic'' scattering processes of the form
$(k_{F}^{\ell};-k_{F}^{-\ell})\rightarrow (k_{F}^{\ell'};-k_{F}^{-\ell'})$ with
$|\ell|\neq|\ell'|$ are also allowed by angular and linear momentum
conservation, but they lead to substantially different and more complicated
effects, in part because they change the numbers of particles in the three
Luttinger liquids. We shall ignore these interactions in the body of this paper,
although a brief discussion of them is located in the Appendix.]

We write the forward scattering terms as generalizations of the interactions
in the Tomonaga-Luttinger model~\cite{ref-tomonaga}:
\begin{equation}
\label{eq-hfs}
H_{{\rm fs}}=\frac{1}{2L}\sum_{q}\left[\begin{array}{c}
\rho^{0}_{+}(q) \\
\rho^{0}_{-}(q) \\
\rho^{+1}_{+}(q) \\
\rho^{+1}_{-}(q) \\
\rho^{-1}_{+}(q) \\
\rho^{-1}_{-}(q)
\end{array}\right]^{T}\left[\begin{array}{cccccc}
g_{00} & g_{00} & g_{01} & g_{01} & g_{01} & g_{01} \\
g_{00} & g_{00} & g_{01} & g_{01} & g_{01} & g_{01} \\
g_{01} & g_{01} & g_{11} & g_{11} & g_{11} & g_{11} \\
g_{01} & g_{01} & g_{11} & g_{11} & g_{11} & g_{11} \\
g_{01} & g_{01} & g_{11} & g_{11} & g_{11} & g_{11} \\
g_{01} & g_{01} & g_{11} & g_{11} & g_{11} & g_{11}
\end{array}\right]\left[\begin{array}{c}
\rho^{0}_{+}(-q) \\
\rho^{0}_{-}(-q) \\
\rho^{+1}_{+}(-q) \\
\rho^{+1}_{-}(-q) \\
\rho^{-1}_{+}(-q) \\
\rho^{-1}_{-}(-q)
\end{array}\right].
\end{equation}
All the interaction strengths $g_{|\ell||\ell'|}$ are presumed to have the same
underlying
origin; however, their values depend upon the radial density profiles
of the two interacting particles, so they depend upon $|\ell|$ and $|\ell'|$.
The $g_{|\ell||\ell'|}$ may most generally be functions of the momentum
transfer $q$, but their large-$q$ behavior is irrelevant to the low-energy
physics, and we shall take the interactions to be momentum-independent.
The symmetries of the matrix in (\ref{eq-hfs}) are physically very reasonable.
The $\ell=0$ states interact symmetrically with the $\ell=\pm1$ states, and the
forward scattering interactions among $|\ell|=1$ states are independent of the
sign of $\ell$.

We shall introduce the remaining Hamiltonian for backward
scattering later, after we look a bit further at the forward-scattering effects.
We may simplify the interactions by introducing symmetric and
antisymmetric combinations of the $|\ell|=1$ operators: $\rho^{S}_{\pm}=
\frac{1}{\sqrt{2}}\left(\rho^{+1}_{\pm}+\rho^{-1}_{\pm}\right)$ and $\rho^{A}
_{\pm}=\frac{1}{\sqrt{2}}\left(\rho^{+1}_{\pm}-\rho^{-1}_{\pm}\right)$.
This change of basis mixes states which are degenerate under $H_{0}$, and it
simplifies $H_{{\rm fs}}$ to
\begin{equation}
H_{{\rm fs}}=\frac{1}{2L}\sum_{q}\left[\begin{array}{c}
\rho^{0}_{+}(q) \\
\rho^{0}_{-}(q) \\
\rho^{S}_{+}(q) \\
\rho^{S}_{-}(q) \\
\rho^{A}_{+}(q) \\
\rho^{A}_{-}(q)
\end{array}\right]^{T}\left[\begin{array}{cccccc}
g_{00} & g_{00} & \sqrt{2}g_{01} & \sqrt{2}g_{01} & 0 & 0 \\
g_{00} & g_{00} & \sqrt{2}g_{01} & \sqrt{2}g_{01} & 0 & 0 \\
\sqrt{2}g_{01} & \sqrt{2}g_{01} & 2g_{11} & 2g_{11} & 0 & 0 \\
\sqrt{2}g_{01} & \sqrt{2}g_{01} & 2g_{11} & 2g_{11} & 0 & 0 \\
0 & 0 & 0 & 0 & 0 & 0 \\
0 & 0 & 0 & 0 & 0 & 0 \\ 
\end{array}\right]\left[\begin{array}{c}
\rho^{0}_{+}(-q) \\
\rho^{0}_{-}(-q) \\
\rho^{S}_{+}(-q) \\
\rho^{S}_{-}(-q) \\
\rho^{A}_{+}(-q) \\
\rho^{A}_{-}(-q)
\end{array}\right].
\end{equation}
The antisymmetric modes are completely decoupled from the others.

The separation off of the antisymmetric $|\ell|=1$ modes is extremely
fortuitous,
because the same separation also occurs when backward scattering is included.
The backscattering Hamiltonian is
\begin{equation}
\label{eq-hbs}
H_{{\rm bs}}=\frac{1}{L}\sum_{k,p,q}\sum_{\ell,\ell'=\pm1}g_{{\rm bs}}
a^{\ell\dag}_{k}b^{\ell'\dag}_{p}a^{\ell'}_{p+2k_{F}+q}b^{\ell}_{p-2k_{F}-q}.
\end{equation}
We have neglected the backscattering for the $\ell=0$ modes, because it may be
rewritten as an effective forward scattering (i.e.\ as a contribution to
$g_{00}$).
$H_{{\rm bs}}$ adds additional terms to the linear coupling matrix and
introduces a sine-Gordon coupling~\cite{ref-schulz,ref-chui,ref-coleman} as well:
\begin{eqnarray}
H_{{\rm fs}+{\rm bs}} & = & \frac{1}{2L}\sum_{q}\left[\begin{array}{c}
\rho^{0}_{+}(q) \\
\rho^{0}_{-}(q) \\
\rho^{S}_{+}(q) \\
\rho^{S}_{-}(q)
\end{array}\right]^{T}\left[\begin{array}{cccc}
g_{00} & g_{00} & \sqrt{2}g_{01} & \sqrt{2}g_{01} \\
g_{00} & g_{00} & \sqrt{2}g_{01} & \sqrt{2}g_{01} \\
\sqrt{2}g_{01} & \sqrt{2}g_{01} & 2g_{11} & 2g_{11}-g_{\rm bs} \\
\sqrt{2}g_{01} & \sqrt{2}g_{01} & 2g_{11}-g_{\rm bs} & 2g_{11}
\end{array}\right]\left[\begin{array}{c}
\rho^{0}_{+}(-q) \\
\rho^{0}_{-}(-q) \\
\rho^{S}_{+}(-q) \\
\rho^{S}_{-}(-q)
\end{array}\right] \nonumber\\
& & -\frac{g_{{\rm bs}}}{L}\sum_{q}\rho^{A}_{+}(q)\rho^{A}_{-}(-q)
+\frac{2g_{{\rm bs}}}{(2\pi\alpha^{1})^{2}}\int\! dx\,\cos\left(\sqrt{8}
\phi^{A}\right).
\end{eqnarray}
$\phi^{A}$ is the boson field associated with the antisymmetric modes,
\begin{equation}
\phi^{A}(x)=-\frac{i\pi}{L}\sum_{p\neq0}\frac{1}{p}e^{-\alpha^{1}|p|/2-ipx}
\left[\rho^{A}_{+}(p)+\rho^{A}_{-}(p)\right].
\end{equation}
The parameter $\alpha^{1}$ is a short-distance cutoff, and $v^{1}_{F}
\alpha^{1}$ is the bandwidth of the $\ell=\pm1$ Luttinger liquids.
(In~\cite{ref-altschul}, we found that $\alpha^{\ell}=1/k^{\ell}_{F}$ for a
system of trapped fermions such as we are considering.)

Much of the remaining work we shall do on this model involves
Bogoliubov transforming the operators in order to simplify the
interactions, replacing the operators $\rho_{\pm}(q)$ with new operators
$\bar{\rho}_{\pm}(q)$ that obey the same commutation
relations~\cite{ref-bogoliubov}. Our transformations are formally unitary
only if the interactions decrease sufficiently quickly as functions of $|q|$.
However, we have previously stated that the large-momentum behavior of the
couplings is irrelevant to the low-energy physics, and for momenta
above $1/\alpha^{\ell}$, the model is unphysical in any case. So we shall
continue to treat the couplings as constants.

Performing an appropriate Bogoliubov transformation on the antisymmetric
modes transforms the corresponding portion of the Hamiltonian into
\begin{equation}
H'=\frac{2\pi u_{A}}{L}\sum_{q>0}
\left[\bar{\rho}^{A}_{+}(q)\bar{\rho}^{A}_{+}(-q)
+\bar{\rho}^{A}_{-}(-q)\bar{\rho}^{A}_{-}(q)\right]+\frac{2g_{{\rm bs}}}
{(2\pi\alpha^{1})^{2}}\int\! dx\,\cos\left[\left(64\frac{2\pi v_{F}^{1}+g
_{{\rm bs}}}{2\pi v_{F}^{1}-g_{{\rm bs}}}\right)^{1/4}\bar{\phi}^{A}\right],
\end{equation}
where the $\bar{\rho}^{A}$ and $\bar{\phi}^{A}$ are the transformed
operators and the corresponding field, and $u_{A}=\sqrt{\left(v_{F}^{1}
\right)^{2}-\left(g_{{\rm bs}}/2\pi\right)^{2}}$. For repulsive interactions
($g_{{\rm bs}}>0$),
the cosine term is renormalized to zero at long wavelengths. For
$g_{{\rm bs}}<0$, the flow is not toward $g_{{\rm bs}}=0$, but for a particular
choice of couplings, the Hamiltonian is exactly solvable~\cite{ref-luther1}.

The self-coupled sine-Gordon field $\bar{\phi}^{A}$ describes one set of
fundamental modes of the system. We now turn our attention to the remaining
modes.
With another Bogoliubov transformation, the remainder of the Hamiltonian
becomes
\begin{equation}
\label{eq-h''}
H''=\frac{2\pi}{L}\sum_{q>0}\left[\begin{array}{c}
\bar{\rho}^{0}_{+}(q) \\
\bar{\rho}^{0}_{-}(q) \\
\bar{\rho}^{S}_{+}(q) \\
\bar{\rho}^{S}_{-}(q)
\end{array}\right]^{T}\left[\begin{array}{cccc}
u_{0} & 0 & \bar{g} & \bar{g}\\
0 & u_{0} & \bar{g} & \bar{g} \\
\bar{g} & \bar{g} & u_{S} & 0 \\
\bar{g} & \bar{g} & 0 & u_{S}
\end{array}\right]\left[\begin{array}{c}
\bar{\rho}^{0}_{+}(-q) \\
\bar{\rho}^{0}_{-}(-q) \\
\bar{\rho}^{S}_{+}(-q) \\
\bar{\rho}^{S}_{-}(-q)
\end{array}\right].
\end{equation}
The transformed coupling is
\begin{equation}
\bar{g}=\left[4\left(\frac{2\pi v_{F}^{0}}{2\pi v_{F}^{0}+2g_{00}}\right)
\left(\frac{2\pi v_{F}^{1}+g_{{\rm bs}}}{2\pi v_{F}^{1}+4g_{11}-g_{{\rm bs}}}
\right)\right]^{1/4}\frac{g_{01}}{2\pi},
\end{equation}
and the wave speeds at $\bar{g}=0$ are
\begin{eqnarray}
u_{0} & = & \sqrt{\left(v_{F}^{0}\right)^{2}+v_{F}^{0}g_{00}/\pi} \\
u_{S} & = & \sqrt{\left(v_{F}^{1}+g_{{\rm bs}}/2\pi\right)\left(v_{F}^{1}+
2g_{11}/\pi-g_{{\rm bs}}/2\pi\right)}
\end{eqnarray}
We shall denote the matrix appearing in (\ref{eq-h''}) by $M$.

If the interactions $g_{00}$, $g_{11}$, and $g_{{\rm bs}}$ conspire
to produce a degeneracy, with $u_{0}=u_{S}=u$, then $M$ may be diagonalized
easily. We again take symmetric and antisymmetric combinations
$\bar{\rho}^{B}_{\pm}=\frac{1}{\sqrt{2}}\left(\bar{\rho}^{0}_{\pm}+\bar{\rho}
^{S}_{\pm}\right)$ and $\bar{\rho}^{C}_{\pm}=\frac{1}{\sqrt{2}}\left(\bar{\rho}
^{0}_{\pm}-\bar{\rho}^{S}_{\pm}\right)$, which transforms $H''$ into the form
\begin{equation}
H''=\frac{2\pi}{L}\sum_{q>0}\left[\begin{array}{c}
\bar{\rho}^{B}_{+}(q) \\
\bar{\rho}^{B}_{-}(q) \\
\bar{\rho}^{C}_{+}(q) \\
\bar{\rho}^{C}_{-}(q)
\end{array}\right]^{T}\left[\begin{array}{cccc}
u+\bar{g} & \bar{g} & 0 & 0 \\
\bar{g} & u+\bar{g} & 0 & 0 \\
0 & 0 & u-\bar{g} & -\bar{g} \\
0 & 0 & -\bar{g} & u-\bar{g}
\end{array}\right]\left[\begin{array}{c}
\bar{\rho}^{B}_{+}(-q) \\
\bar{\rho}^{B}_{-}(-q) \\
\bar{\rho}^{C}_{+}(-q) \\
\bar{\rho}^{C}_{-}(-q)
\end{array}\right].
\end{equation}
One further Bogliubov transformtion gives the decoupled normal modes, with
frequencies $\omega=|q|\sqrt{u^{2}\pm 2u\bar{g}}$.

In the presence of a $u_{0}=u_{S}$ degeneracy, we find exact energy shifts
containing terms linear in $\bar{g}$. In the absence of this special
degeneracy, we expect the lowest-order correction to the frequency to be
${\cal O}(\bar{g}^{2})$. We may find this leading term by treating $\bar{g}$
perturbatively. Through a completely
straightforward calculation, we find shifted frequencies
\begin{eqnarray}
\omega_{0} & \approx & |q|u_{0}+|q|\bar{g}^{2}\frac{2u_{S}}{u_{0}^{2}-u_{S}
^{2}} \\
\omega_{S} & \approx & |q|u_{S}-|q|\bar{g}^{2}\frac{2u_{0}}{u_{0}^{2}-u_{S}
^{2}}
\end{eqnarray}
for the $\bar{\rho}^{0}$ and  $\bar{\rho}^{S}$ modes, respectively.

If we wish to write $H''$ (for general $u_{0}$, $u_{S}$, and $\bar{g}$) as a
sum of noninteracting harmonic oscillator Hamiltonians, we
must find a $U(2,2)$ matrix that diagonalizes $M$. That is, we need to
find a matrix $V$ such that $V^{\dag}MV$ is diagonal, and $V^{\dag}I_{2,2}V=
I_{2,2}$, where $I_{2,2}=$ diag [1 $-1$ 1 $-1$]. [Our subsidiary condition
on $V$---which makes it a member of $U(2,2)$, by definition---ensures that the
transformation
preserves the correct commutation relations. The Bogliubov transformation that
led to the form (\ref{eq-h''}) for $H''$ corresponds to an element of the block
diagonal $U(1,1)\times U(1,1)$ subgroup of $U(2,2)$.]

The $U(2,2)$ diagonalization problem for $M$ is equivalent to the ordinary
diagonalization of
\begin{equation}
M'=\left[\begin{array}{cccc}
u_{0} & 0 & \bar{g} & i\bar{g}\\
0 & -u_{0} & i\bar{g} & -\bar{g} \\
\bar{g} & i\bar{g} & u_{S} & 0 \\
i\bar{g} & -\bar{g} & 0 & -u_{S}
\end{array}\right],
\end{equation}
which is related to $M$ by a phase rotation of the $\bar{\rho}_{-}$
terms. The positive eigenvalues of $M'$ are exactly the normal mode
frequencies, and the eigenvectors of $M'$ give (upon an appropriate undoing of
the phase rotation) the weights of the various $\bar{\rho}$ in the normal
mode creation and annihilation operators.

The positive eigenvalues of $M'$ give normal mode frequencies of
\begin{equation}
\label{eq-omega}
\omega=\frac{|q|}{\sqrt{2}}\sqrt{u_{0}^{2}+u_{S}^{2}\pm\sqrt{\left(u_{0}^{2}-
u_{S}^{2}\right)^{2}+16\bar{g}^{2}u_{0}u_{S}}}.
\end{equation}
When $u_{0}=u_{S}$, we recover our earlier result, and for $u_{0}\neq u_{S}$,
the exact expression reduces to the perturbative result up to ${\cal O}(\bar{g}
^{2})$.  The corresponding eigenvectors of $M'$ are fairly complicated, but
they may be found by elementary means.

This gives a complete description of the modes of the three interacting
Luttinger liquids. Two set of modes---which are linear combinations of
$\rho^{0}$ and
$\rho^{S}$---describe linear harmonic fluids, with the two different wave
speeds given by (\ref{eq-omega}). The remaining modes are coupled through
a sine-Gordon interaction~\cite{ref-coleman,ref-rajaraman}. If we fix the
number of particles having each value of $\ell$, then, under the
approximations of the
Luttinger model, the $\rho^{\ell}$ operators generate a complete set of states,
and the modes we have found are the only low-energy excitations for the system.
Because the interactions $H_{{\rm fs}}$ and $H_{{\rm bs}}$
are fairly general and physically quite reasonable, the spectrum we have
derived should be highly relevant to the study of trapped fermions in
situations where Luttinger liquid behavior is observable.

\section*{Acknowledgments}
The author is grateful to K. Huang for many helpful discussions.
This work is supported in part by funds provided by the U. S.
Department of Energy (D.O.E.) under cooperative research agreement
DE-FG02-91ER40661.

\appendix

\section*{Appendix: Hamiltonian for ``inelastic'' scattering}

In the body of this paper, we have considered only those scattering processes
which preserve
the individuals particles' $\ell$ values. However, there exists another type
of scattering interaction, which does not have this property. Interactions of
the type $(k_{F}^{\ell};-k_{F}^{-\ell})\rightarrow (k_{F}^{\ell'};
-k_{F}^{-\ell'})$ with $|\ell|\neq|\ell'|$ are also allowed by linear and
angular momentum
conservation. In this appendix, we shall examine the effects of such terms.
We shall find that they lead to interactions that are similar to
sine-Gordon interactions but more complicated.

We begin with an interaction Hamiltonian that describes the scattering
process in question:
\begin{equation}
H_{{\rm in}}=\frac{1}{L}\sum_{k,p,q}\sum_{\ell=\pm1}g_{{\rm in}}a_{k}^{\ell}
b_{-k+q}^{-\ell}a_{p}^{0\dag}b_{-p+q}^{0\dag}+{\rm h.c.},
\end{equation}
where ``h.c.'' denotes the Hermitian conjugate. We may simplify this by
Fourier transforming the $a$ and $b$ operators and expanding the resultant
fermion operators in terms of the boson field. For example, $a_{k}^{+1}$
may be written
as~\cite{ref-mattis2,ref-luther2,ref-heidenreich,ref-haldane1,ref-haldane2}
\begin{eqnarray}
a_{k}^{+1} & = & \frac{1}{\sqrt{L}}\int_{0}^{L}dx\, e^{-ikx}\psi_{+}^{+1}
(x) \nonumber\\
& = & \frac{1}{\sqrt{2\pi\alpha^{1}L}}U^{+1}_{+}\int_{0}^{L}dx\, e^{-ikx}
e^{ik^{1}_{F}x}e^{-i\phi^{+1}(x)+i\theta^{+1}(x)}.
\end{eqnarray}
The operator $U^{+1}_{+}$ carries the fermion number of $a_{k}^{+1}$ but does
not affect the bosonic state of the system, while $\phi^{+1}$ is the boson
field corresponding to the $\ell=+1$ fermion states, and $\theta^{+1}$ is the
integral of the momentum conjugate to $\phi^{+1}$. In general, the operators
$\phi^{\ell}$ and $\theta^{\ell}$ have the forms
\begin{eqnarray}
-i\phi^{\ell} & = & -\frac{\pi}{L}\sum_{p\neq0}\frac{1}{p}e^{-\alpha|p|/2}
e^{-ipx}[\rho^{\ell}_{+}(p)+\rho^{\ell}_{-}(p)]+i\pi\frac{N^{\ell}_{+}+N^{\ell}
_{-}}{L}x \\
i\theta^{\ell} & = & -\frac{\pi}{L}\sum_{p\neq0}\frac{1}{p}e^{-\alpha|p|/2}
e^{-ipx}[\rho^{\ell}_{+}(p)-\rho^{\ell}_{-}(p)]+i\pi\frac{N^{\ell}_{+}-N^{\ell}
_{-}}{L}x,
\end{eqnarray}
where $N^{\ell}_{+}$ and $N^{\ell}_{-}$ are the numbers of excess right- and
left-moving fermions of angular momentum $\ell$, respectively.

If we express all the operators in $H_{{\rm in}}$ in this (bosonized)
fashion, we get
\begin{eqnarray}
H_{{\rm in}} & = & \frac{g_{{\rm in}}}{(2\pi)^{2}\alpha^{1}\alpha^{0}L^{3}}
\sum_{k,p,q}\sum_{\ell=\pm1}\int_{0}^{L}dx_{1}\,dx_{2}\,dx_{3}\,dx_{4}\,
e^{ik(x_{2}-x_{1})}e^{-ik_{F}^{\ell}(x_{2}-x_{1})}e^{iq(x_{4}-x_{2})}
e^{ip(x_{3}-x_{4})} \nonumber\\
& & e^{-ik_{F}^{0}(x_{3}-x_{4})}e^{-i\phi^{\ell}(x_{1})+i\theta^{\ell}(x_{1})}
e^{i\phi^{-\ell}(x_{2})+i\theta^{-\ell}(x_{2})}e^{i\phi^{0}(x_{3})
-i\theta^{0}(x_{3})}e^{-i\phi^{0}(x_{4})-i\theta^{0}(x_{4})}+{\rm h.c.}
\end{eqnarray}
The sums over $k$, $p$, and $q$ give $\delta$-functions according to
$\sum_{k}e^{ikx}
=L\delta(x)$, and these simplify the expression substantially. We find that
\begin{equation}
H_{{\rm in}}=\frac{g_{{\rm in}}}{(2\pi)^{2}\alpha^{0}\alpha^{1}}\sum_{\ell=
\pm1}\int_{0}^{L}dx\,e^{-i\phi^{\ell}(x)+i\theta^{\ell}(x)}e^{i\phi^{-\ell}(x)+
i\theta^{-\ell}(x)}e^{-2i\theta^{0}(x)}+{\rm h.c.}
\end{equation}
Since $\phi^{\ell}$ and $\theta^{\ell}$ are Hermitian, this is equivalent to
\begin{eqnarray}
H_{{\rm in}} & = & \frac{2g_{{\rm in}}}{(2\pi)^{2}\alpha^{0}\alpha^{1}}
\sum_{\ell=\pm1}\int_{0}^{L}dx\, \cos\left[\phi^{\ell}(x)-\theta^{\ell}(x)-
\phi^{-\ell}(x)-\theta^{-\ell}(x)+2\theta^{0}(x)\right] \nonumber\\
& = & \frac{g_{{\rm in}}}{\pi^{2}\alpha^{1}\alpha^{0}}\int_{0}^{L}dx\,
\cos\left[\phi^{+1}(x)-\phi^{-1}(x)\right]\cos\left[\theta^{+1}(x)+\theta
^{-1}(x)-2\theta^{0}(x)\right].
\end{eqnarray}
This a new nonlinear interaction. It is similar to a
sine-Gordon interaction, but it contains substantial additional complexity.
The effects of this interaction would presumably need to be determined
perturbatively.

\end{document}